\documentclass[journal,twoside,web]{ieeecolor}

\usepackage{lcsys}

\usepackage{url}
\usepackage{graphicx,graphics,epsfig,epstopdf}
\usepackage[utf8]{inputenc} 
\usepackage[T1]{fontenc}    
\usepackage{url}            
\usepackage{booktabs}       

\usepackage{amsfonts}       
\usepackage{amssymb,amsmath,amsthm}
\usepackage{nicefrac}       
\usepackage{microtype}      
\usepackage{tikz}
\usepackage{caption}
\usepackage{wrapfig}
\usepackage[linesnumbered,ruled,vlined]{algorithm2e}
\usepackage{subcaption}
\usepackage{soul}
\usepackage{mathtools}
\usepackage{accents}
\usepackage{setspace}
\usepackage{multirow,makecell}
\usepackage{threeparttable}
\usepackage{optidef}
\usepackage{multirow}
\usepackage{multicol}
\usepackage{makecell}
\usepackage{array} 
\usepackage{cite}
\usepackage{wrapfig}

\SetKwInput{kwAdd}{Add}
\SetKwInput{kwSet}{Set}
\SetKwInput{kwReset}{Reset}
\SetKwInput{kwUpdate}{Update}
\SetKwInput{kwClear}{Clear}
\SetKwInput{kwSave}{Save}
\SetKwInput{kwBuild}{Build}
\SetKwInput{kwIni}{Initialization}
\usetikzlibrary{matrix,chains,positioning,decorations.pathreplacing,arrows}

\newtheorem{lemma}{Lemma}
\newtheorem{remark}{Remark}
\newtheorem{theorem}{Theorem}

\newcommand{\col}{{\rm col\;}}

\newtheorem{alemma}{Lemma}[section]

\begin{document}


\allowdisplaybreaks
\title{Safe Online Control-Informed Learning}

\author{
Tianyu Zhou$^{1}$, Zihao Liang$^{2}$, Zehui Lu$^{1}$ and Shaoshuai Mou$^{1}$
\thanks{$^{1}$T. Zhou, Z. Lu and S. Mou are with the School of Aeronautics and Astronautics, Purdue University, IN 47907, USA {\tt\small \{zhou1043, lu846, mous\}@purdue.edu}}
\thanks{$^{2}$Z. Liang is an independent researcher {\tt\small danlzh@outlook.com}}
}

\maketitle
\thispagestyle{empty}

\begin{abstract}

This paper proposes a Safe Online Control-Informed Learning framework for safety-critical autonomous systems. The framework unifies optimal control, parameter estimation, and safety constraints into an online learning process. It employs an extended Kalman filter to incrementally update system parameters in real time, enabling robust and data-efficient adaptation under uncertainty. A softplus barrier function enforces constraint satisfaction during learning and control while eliminating the dependence on high-quality initial guesses. Theoretical analysis establishes convergence and safety guarantees, and the framework's effectiveness is demonstrated on cart-pole and robot-arm systems.

\end{abstract}


\IEEEpeerreviewmaketitle

\section{Introduction}


\IEEEPARstart{I}{nformed} Machine Learning (IML), such as physics-informed learning, integrates prior knowledge such as physical laws, expert insights, or existing models into machine learning (ML) to enable reliable and interpretable predictions in noisy or data scarce settings \cite{karniadakis2021physics}. Extending this idea, Control Informed Learning (CIL) unifies control theory and ML for autonomous systems. Control theory provides principled models and optimization, while ML offers adaptability. This synergy reduces complexity, enhances applicability, and ensures learning remains consistent with established models, improving both reliability and accuracy \cite{jin2020pontryagin,liang2025online,jin2022directional,zhang2025constrained}. Several studies apply CIL by embedding optimal control (OC) principles into learning. Early work used implicit planners and unrolled computational graphs, casting OC problems as repeated gradient descent \cite{pereira2018mpc,srinivas2018universal}. Although effective, this approach is memory-intensive and sensitive to the number of unrolled steps. Differentiable MPC eased these issues with a Linear Quadratic Regulator (LQR) approximation but still required expensive matrix inversions \cite{amos2018differentiable}. Pontryagin’s Differential Programming (PDP) later reduced memory and computation by differentiating through Pontryagin’s Maximum Principle \cite{jin2020pontryagin}, yet remained limited by batch gradient descent and its inability to handle noise. For autonomous systems, online adaptation is crucial. The Online Control Informed Learning (OCIL) framework addresses this by combining OC with state estimation to improve data efficiency, handle noise, and enable real-time updates \cite{liang2025online}. Using an extended Kalman filter, OCIL updates parameters on the fly, enhancing robustness though not explicitly ensuring safety.

For autonomous systems, safety is defined by constraints on states and inputs that must hold during learning and execution; violations can cause irreversible failures, making such systems safety-critical \cite{nilim2005robust,ames2016control}. Barrier functions, such as logarithmic or ReLU penalties, are often used to convert constrained problems into unconstrained ones, enabling gradient-based optimization \cite{garcia2015comprehensive,jin2020neural}. Building on this idea, Safe Pontryagin Differentiable Programming (Safe PDP) provides a differentiable framework that enforces safety by computing gradients through an unconstrained auxiliary system \cite{jin2021safe}. However, Safe PDP is not an online method as it relies on batch optimization through repeated gradient descent, making it computationally expensive and unsuitable for real-time adaptation. It also requires a feasible initialization to ensure the log-barrier loss is well-defined; to mitigate this, \cite{jin2021safe} proposed initializing with supervised learning from safe demonstrations. Moreover, Safe PDP does not explicitly handle noisy measurements. These limitations, including its batch nature, dependence on demonstrations, and sensitivity to noise, restrict its applicability.

To address these limitations, this paper proposes a Safe OCIL framework. The contributions of this work are threefold: 1) Safe OCIL incorporates an online parameter estimation scheme that incrementally tunes the unknown parameters through state estimation, enabling real-time adaptation to dynamic environments and noisy data. 2) We establish theoretical safety guarantees by ensuring constraint satisfaction throughout both the learning and control processes. 3) We design a modified barrier function to address the limits of existing ones: logarithmic barriers need a feasible starting point, and ReLU functions are not differentiable. Our approach uses a smoothed version of ReLU that is both initialization-free and differentiable, which avoids solver failures and ensures feasibility and safety during learning.


\noindent\textbf{Notations.} 
$\lVert \cdot \lVert$ denotes the Euclidean norm.
$\text{col}\{\boldsymbol{v}_1,\hdots,\boldsymbol{v}_a\}$ denotes a column stack of elements $\boldsymbol{v}_1,\hdots,\boldsymbol{v}_a$.
\section{Problem Formulation}\label{ProF}

Consider the OC system $\boldsymbol{\Sigma}(\boldsymbol{\theta})$, with parameter $\boldsymbol{\theta}\in\mathbb{R}^p$, whose behavior is defined by minimizing a control objective:
{\small
\begin{subequations} \label{eq:optiOC}
\begin{align}
&\{\boldsymbol{x}_{1:T}(\boldsymbol{\theta}),\boldsymbol{u}_{0:T-1}(\boldsymbol{\theta})\} = \arg \min_{\boldsymbol{x}_{1:T},\boldsymbol{u}_{0:T-1}} J(\boldsymbol{\theta}), \label{eq:tru_obj}\\
\text{s.t. } &\boldsymbol{x}_{t+1} = \boldsymbol{f}(\boldsymbol{x}_t, \boldsymbol{u}_t, \boldsymbol{\theta}), \  \text{with}\ \boldsymbol{x}_0\ \text{given}, \label{eq:tru_dyn} \\
&\boldsymbol{g}_t (\boldsymbol{x}_t,\boldsymbol{u}_t,\boldsymbol{\theta})\leq \boldsymbol{0},
\boldsymbol{h}_t (\boldsymbol{x}_t,\boldsymbol{u}_t,\boldsymbol{\theta}) = \boldsymbol{0}, \forall t=0,\cdots,T-1, \label{eq:path_const}\\
&\boldsymbol{g}_T (\boldsymbol{x}_T,\boldsymbol{\theta})\leq \boldsymbol{0}, \boldsymbol{h}_T (\boldsymbol{x}_T,\boldsymbol{\theta})= \boldsymbol{0},\label{eq:final_const}
\end{align}  
\end{subequations}
}%
where $J(\boldsymbol{\theta})= \textstyle\sum\nolimits_{t=0}^{T-1}c_t (\boldsymbol{x}_t,\boldsymbol{u}_t,\boldsymbol{\theta})+c_T (\boldsymbol{x}_T,\boldsymbol{\theta})$, $t=0,1,2,\cdots, T$ is the time index with $T$ being the final time; $\boldsymbol{x}_t\in\mathbb{R}^n$ and $\boldsymbol{u}_t \in\mathbb{R}^m$ denote the system state and control input, respectively;  
$\boldsymbol{x}_{0:T}(\boldsymbol{\theta})\triangleq\col\{\boldsymbol{x}_0(\boldsymbol{\theta}),\dots,\boldsymbol{x}_T(\boldsymbol{\theta})\}$ 
and $\boldsymbol{u}_{0:T-1}(\boldsymbol{\theta})\triangleq\col\{\boldsymbol{u}_0(\boldsymbol{\theta}),\dots,\boldsymbol{u}_{T-1}(\boldsymbol{\theta})\}$ denote the state and input trajectories for parameter $\boldsymbol{\theta}$, respectively;
$\boldsymbol{f}:\mathbb{R}^n\times\mathbb{R}^m\times\mathbb{R}^p\rightarrow\mathbb{R}^n$ denotes a twice-differentiable time-invariant system dynamics; $c_t:\mathbb{R}^n\times\mathbb{R}^m\times\mathbb{R}^p\mapsto\mathbb{R}$ and $c_T:\mathbb{R}^n\times\mathbb{R}^p\mapsto\mathbb{R}$ denote running cost the final cost, respectively; $\boldsymbol{g}_t:\mathbb{R}^n \times \mathbb{R}^m \times \mathbb{R}^p \rightarrow\mathbb{R}^{q_t}$ and $\boldsymbol{h}_t:\mathbb{R}^n \times \mathbb{R}^m \times \mathbb{R}^p \rightarrow\mathbb{R}^{s_t}$ denote the immediate inequality and equality constraints at time $t$, respectively; $\boldsymbol{g}_T:\mathbb{R}^n \times \mathbb{R}^p \rightarrow\mathbb{R}^{q_T}$ and $\boldsymbol{h}_T:\mathbb{R}^n \times \mathbb{R}^p \rightarrow\mathbb{R}^{s_T}$ denote the terminal inequality and equality constraints, respectively. All cost functions and constraints are assumed twice-differentiable.
For notation simplicity, we define the trajectory of the OC system $\boldsymbol{\Sigma}(\boldsymbol{\theta})$ as $\boldsymbol{\xi}({\boldsymbol{\theta}})\triangleq
\text{col}\{ \boldsymbol{x}_{0:T}(\boldsymbol{\theta}),
\boldsymbol{u}_{0:T-1}(\boldsymbol{\theta}) \}
\in\mathbb{R}^{(T+1)n+Tm}.$


A demonstration trajectory $\boldsymbol{\xi}^*$ is generated from a demo system $\boldsymbol{\Sigma}(\boldsymbol{\theta}^*)$, where $\boldsymbol{\theta}^*$ is the true parameters and remain unchanged. The trajectory generated from \eqref{eq:optiOC} matches the demonstration trajectory when $\boldsymbol{\theta}=\boldsymbol{\theta}^*$, i.e., $\boldsymbol{\xi}(\boldsymbol{\theta}^*)=\boldsymbol{\xi}^*$.
At each time $t$, a noisy measurement is observed:
{\small
\begin{equation}
    \boldsymbol{y}_t^* =\boldsymbol{z}(\boldsymbol{\xi}_t^*)+\boldsymbol{v}_t\in \mathbb{R}^r,
\end{equation}
}%
where $\boldsymbol{z}:\mathbb{R}^{n+m}\rightarrow\mathbb{R}^r$ denotes a twice-differentiable measurement function; $\boldsymbol{v}_t\sim\mathcal{N}(\boldsymbol{0}_r,\boldsymbol{R}_t)$ denotes a Gaussian measurement noise at time $t$; and $\boldsymbol{R}_t\in\mathbb{R}^{r\times r}$ denotes the covariance matrices of the noise. 
We define a stage loss: 
{\small
\begin{equation}
    \boldsymbol{l}(\boldsymbol{\xi}_t(\boldsymbol{\theta}),\boldsymbol{y}_t^*)=\boldsymbol{y}_t^*-\boldsymbol{z}(\boldsymbol{\xi}_t(\boldsymbol{\theta}))\in\mathbb{R}^r.
\end{equation}
}%
The performance of the entire trajectory can be evaluated by a cumulative loss, which is assumed to be twice-differentiable: 
{\small
\begin{equation}\label{eq:L}
    L(\boldsymbol{\xi}(\boldsymbol{\theta}))=\textstyle\sum_{t=0}^{T}||\boldsymbol{l}(\boldsymbol{\xi}_t(\boldsymbol{\theta}),\boldsymbol{y}_t^*)||^2.
\end{equation}
}%
Define $\hat{\boldsymbol{\theta}}_t\in\mathbb{R}^p$ as an estimation of the true parameter $\boldsymbol{\theta}^*$ at time $t$. 
The \textit{problem of interest} is to design an online update law for $\hat{\boldsymbol{\theta}}_t$ based on the observations $\boldsymbol{y}_t^*$, such that $\lim_{t \to \infty} L(\boldsymbol{\xi}(\hat{\boldsymbol{\theta}}_t)) \rightarrow 0$.



\section{Methodology}\label{sec:method}

In this section, we introduce the proposed Safe OCIL algorithm. The approach begins with an online parameter estimator inherited from OCIL, which continuously updates unknown system parameters in real time. Next, we analyze the differentiability of $\boldsymbol{\xi}(\boldsymbol{\theta})$, a property that enables efficient gradient-based updates. We then approximate the gradient of the trajectory through a safe unconstrained system, thereby avoiding the computational burden of directly solving constrained OC problems while still providing safety guarantees. Finally, we establish through theoretical analysis that the proposed algorithm achieves both convergence and safety.

\subsection{Online Parameter Estimator}


Inspired by \cite{liang2025online}, the parameter optimization problem is converted into a state estimation problem, where the unknown $\boldsymbol{\theta}^*$ is treated as the state to be estimated. 
The online update is performed using the following update laws:
{\small
\begin{subequations}\label{eq:EKF}
\begin{align}
    \text{Predict: }& \hat{\boldsymbol{\theta}}_{t|t-1} := \hat{\boldsymbol{\theta}}_{t-1}, \quad \boldsymbol{P}_{t|t-1} := \boldsymbol{P}_{t-1},\label{eq:predict}\\
    \text{Update: }& \boldsymbol{K}_t = \boldsymbol{P}_{t|t-1} \boldsymbol{L}_t^\top (\boldsymbol{L}_t \boldsymbol{P}_{t|t-1} \boldsymbol{L}_t^\top + \boldsymbol{R}_t)^{-1},\label{eq:H_t}\\
    & \boldsymbol{P}_t = (\boldsymbol{I}-\boldsymbol{K}_t \boldsymbol{L}_t)\boldsymbol{P}_{t|t-1},\\
    & \hat{\boldsymbol{\theta}}_t = \hat{\boldsymbol{\theta}}_{t|t-1} - \boldsymbol{K}_t (\boldsymbol{y}_t^* - \boldsymbol{z}(\hat{\boldsymbol{\theta}}_{t|t-1})),\label{eq:theta_update}
\end{align}
\end{subequations}
}%
where
$\boldsymbol{L}_t = \frac{d \boldsymbol{l}(\boldsymbol{\xi}_t(\boldsymbol{\theta}_{t}),\boldsymbol{y}_t^*)}{d \boldsymbol{\theta}_{t}}|_{\boldsymbol{\theta}_t=\hat{\boldsymbol{\theta}}_{t|t-1}}\in\mathbb{R}^{r \times p}$. 
The subscript $_{t|t-1}$ indicates prediction prior to incorporating the latest measurement. The positive definite matrix $\boldsymbol{P}\in\mathbb{R}^{p\times p}$
denotes the covariance of the estimate, $\boldsymbol{K}_t\in\mathbb{R}^{p\times r}$ denotes the Kalman gain, and $\boldsymbol{I}$ denotes the identity matrix. 
\begin{lemma}\label{lemma:OCIL}
\textbf{(Online Parameter Estimator \cite{liang2025online})}. Define the estimation error $\tilde{\boldsymbol{\theta}}_t=\boldsymbol{\theta}^*-\hat{\boldsymbol{\theta}}_t$, measurement error $\boldsymbol{e}_t=\boldsymbol{l}(\boldsymbol{\xi}(\boldsymbol{\theta}^*))-\boldsymbol{l}(\boldsymbol{\xi}(\hat{\boldsymbol{\theta}}_{t|t-1}))$, and prediction error $\tilde{\boldsymbol{\theta}}_{t|t-1}=\boldsymbol{\theta}^*-\hat{\boldsymbol{\theta}}_{t|t-1}$.  
Assume $\boldsymbol{L}_t$ is full rank and uniformly bounded. 
With the diagonal matrices $\boldsymbol{\mathcal{F}}_t \in \mathbb{R}^{r \times r}$ and $\boldsymbol{\mathcal{G}}_t \in \mathbb{R}^{p \times p}$ introduced by \cite{boutayeb2002convergence}, if $(\boldsymbol{\mathcal{F}}_t-\boldsymbol{I})^2 \leq \boldsymbol{R}_t (\boldsymbol{L}_t \boldsymbol{P}_{t|t-1} \boldsymbol{L}_t' + \boldsymbol{R}_t)^{-1}, \boldsymbol{\mathcal{G}}_t' \boldsymbol{P}_t^{-1}\boldsymbol{\mathcal{G}}_t -\boldsymbol{P}_t^{-1} \leq 0,$
then the estimator \eqref{eq:EKF} ensures local asymptotic convergence, i.e. $\lim_{t\rightarrow\infty} \tilde{\boldsymbol{\theta}}_t=\boldsymbol{0}$.
\end{lemma}

The only unknown term in \eqref{eq:EKF} is $\boldsymbol{L}_t$, which is challenging to compute due to the implicit dependence of the stage loss $\boldsymbol{l}(\boldsymbol{\xi}_t(\boldsymbol{\theta}),\boldsymbol{y}_t^*)$ on $\boldsymbol{\theta}$. 
For notation simplicity, $\frac{d \boldsymbol{l}(\boldsymbol{\xi}_t(\boldsymbol{\theta}_{t}),\boldsymbol{y}_t^*)}{d \boldsymbol{\theta}_{t}}|_{\boldsymbol{\theta}_t=\hat{\boldsymbol{\theta}}_{t|t-1}}$ is written as $\frac{d \boldsymbol{l}(\boldsymbol{\xi}_t(\boldsymbol{\theta}))}{d\boldsymbol{\theta}}$.
To obtain this gradient, one can employ the chain rule
{\small
\begin{equation}\label{eq:chain}
    \frac{d \boldsymbol{l}(\boldsymbol{\xi}_t(\boldsymbol{\theta}))}{d\boldsymbol{\theta}} = \frac{\partial \boldsymbol{l}(\boldsymbol{\xi}_t(\boldsymbol{\theta}))}{\partial \boldsymbol{\xi}_t(\boldsymbol{\theta})} \frac{\partial\boldsymbol{\xi}_t(\boldsymbol{\theta})}{\partial \boldsymbol{\theta}},
\end{equation}
}%
where
$\frac{\partial \boldsymbol{l}(\boldsymbol{\xi}_t(\boldsymbol{\theta}))}{\partial \boldsymbol{\xi}_t(\boldsymbol{\theta})}$ 
is known as the stage loss function is known. The remaining challenge is to find $\frac{\partial\boldsymbol{\xi}_t(\boldsymbol{\theta})}{\partial \boldsymbol{\theta}}$.

\subsection{Differentiability of $\boldsymbol{\xi}(\boldsymbol{\theta})$}
We define the following Hamiltonian for $\boldsymbol{\Sigma}(\boldsymbol{\theta})$ in \eqref{eq:optiOC}:
{\small
\begin{subequations}\label{eq:CHamiltonian}
\begin{align}
    \boldsymbol{\mathcal{L}}_t = & c_t(\boldsymbol{x}_t,\boldsymbol{u}_t,\boldsymbol{\theta}) + \boldsymbol{\lambda}_{t+1}' \boldsymbol{f}(\boldsymbol{x}_t,\boldsymbol{u}_t,\boldsymbol{\theta}) \\
    & + \boldsymbol{\mu}_t' \boldsymbol{g}_t(\boldsymbol{x}_t,\boldsymbol{u}_t,\boldsymbol{\theta}) + \boldsymbol{\nu}_t'\boldsymbol{h}_t(\boldsymbol{x}_t,\boldsymbol{u}_t,\boldsymbol{\theta}), \nonumber \\
    \boldsymbol{\mathcal{L}}_T = & c_T(\boldsymbol{x}_T,\boldsymbol{\theta}) + \boldsymbol{\mu}_T'\boldsymbol{g}_T(\boldsymbol{x}_T,\boldsymbol{\theta}) + \boldsymbol{\nu}_T'\boldsymbol{h}_T(\boldsymbol{x}_T,\boldsymbol{\theta}).
\end{align}
\end{subequations}
}%
$\boldsymbol{\lambda}_t$ is the costate, $\boldsymbol{\mu}_t\in\mathbb{R}^{q_t}$ and $\boldsymbol{\nu}_t\in\mathbb{R}^{s_t}$ are multipliers for the inequality and equality constraints, respectively. 
The well-known second-order condition for $\boldsymbol{\xi}(\boldsymbol{\theta})$ to be a local isolated minimizing trajectory to $\boldsymbol{\Sigma}(\boldsymbol{\theta})$ is well-established in \cite{pearson1966discrete}. 
With this condition, \cite{jin2021safe} has the following result:
\begin{lemma}\label{lemma:Differentiability}
\textbf{(Differentiability of $\boldsymbol{\xi}(\boldsymbol{\theta})$ \cite{jin2021safe})}.  
Given fixed $\bar{\boldsymbol{\theta}}$, assume (i) the second-order condition \eqref{eq:second_order} holds for $\boldsymbol{\Sigma}(\bar{\boldsymbol{\theta}})$; 
(ii) the gradients of all active constraints at $\boldsymbol{\xi}(\bar{\boldsymbol{\theta}})$ are linearly independent; and 
(iii) complementarity holds with positive multipliers. 
Then, for all $\boldsymbol{\theta}$ in a neighborhood of $\bar{\boldsymbol{\theta}}$, there exists a unique $C^1$ function 
$(\boldsymbol{\xi}(\boldsymbol{\theta}),\boldsymbol{\lambda}_{1:T}(\boldsymbol{\theta}), \boldsymbol{\mu}_{0:T}(\boldsymbol{\theta}),\boldsymbol{\nu}_{0:T}(\boldsymbol{\theta}))$ 
satisfying these conditions for the constrained system $\boldsymbol{\Sigma}(\boldsymbol{\theta})$ and coinciding with 
$(\boldsymbol{\xi}(\bar{\boldsymbol{\theta}}),\boldsymbol{\lambda}_{1:T}(\bar{\boldsymbol{\theta}}), \boldsymbol{\mu}_{0:T}(\bar{\boldsymbol{\theta}}),\boldsymbol{\nu}_{0:T}(\bar{\boldsymbol{\theta}}))$ at $\boldsymbol{\theta}=\bar{\boldsymbol{\theta}}$. 
Hence, $\boldsymbol{\xi}(\boldsymbol{\theta})$ is a local isolated minimizing trajectory of $\boldsymbol{\Sigma}(\boldsymbol{\theta})$.


\end{lemma}
With Lemma \ref{lemma:Differentiability}, \cite{jin2021safe} achieves $\frac{\partial \boldsymbol{\xi}_t(\boldsymbol{\theta})}{\partial \boldsymbol{\theta}}$ via an auxiliary control system using an equality-constrained LQR. Although theoretically sound, this approach faces practical issues, such as solving the constrained problem \eqref{eq:optiOC} being more challenging than the unconstrained case, 
and identifying active constraints being error-sensitive. These motivate converting \eqref{eq:optiOC} into an unconstrained system and approximating the gradient.

\subsection{Gradient Approximator}

Barriers are often used to convert constrained problems into unconstrained ones, improving feasibility and robustness in safety-critical systems. For example, \cite{jin2021safe} applies a logarithmic barrier to enforce inequalities. This keeps trajectories safe but fails if the initial guess violates constraints, potentially causing solver failure. To address this limitation, we use a differentiable approximation of ReLU (softplus function):
{\small
\begin{equation}\label{eq:softplus}
    \phi_{\beta}(x) = \beta\ln (1+e^{x / \beta}),
\end{equation}
}%
where the known parameter $\beta>0$ controls the sharpness of the approximation. This function retains differentiability while penalizing constraint violations, making the optimization more robust to infeasible initial guesses. 
\begin{remark}
    \cite{jin2021safe} requires the initial trajectory to be feasible and generated via supervised learning. 
    To improve computational efficiency, the proposed barrier function \eqref{eq:softplus} is essential for handling initially infeasible guesses while guiding the optimization toward feasibility.
\end{remark}

We convert the constrained system $\boldsymbol{\Sigma}(\boldsymbol{\theta})$ into an \textit{unconstrained system} $\boldsymbol{\Sigma}(\boldsymbol{\theta}, \alpha, \beta)$ by introducing quadratic barrier functions for the equality constraints and softplus barrier functions for the inequality constraints. The trajectory of $\boldsymbol{\Sigma}(\boldsymbol{\theta}, \alpha, \beta)$, $\boldsymbol{\xi}(\boldsymbol{\theta},\alpha,\beta)=\{\boldsymbol{x}_{1:T}(\boldsymbol{\theta},\alpha,\beta),\boldsymbol{u}_{0:T-1}(\boldsymbol{\theta},\alpha,\beta)\}$, is achieved by solving the following OC problem:
\begin{equation}\label{eq:unconstrain_optiOC}
\min_{\boldsymbol{\xi}(\boldsymbol{\theta},\alpha,\beta)} \bar{J}(\boldsymbol{\theta},\alpha,\beta)
\  \text{s.t. } \boldsymbol{x}_{t+1}=\boldsymbol{f}(\boldsymbol{x}_t,\boldsymbol{u}_t,\boldsymbol{\theta}),\; \boldsymbol{x}_0\ \text{given},
\end{equation}
where $\alpha>0$ is the barrier parameter, the augmented objective
{\small
\begin{align}\label{eq:unconstrain}
&\bar{J}(\boldsymbol{\theta}, \alpha, \beta) = 
 \sum_{t=0}^{T-1} ( 
c_t(\boldsymbol{x}_t, \boldsymbol{u}_t, \boldsymbol{\theta}) 
+ \frac{1}{\alpha }\sum_{i=1}^{q_t} \phi_{\beta}(g_{t,i}(\boldsymbol{x}_t, \boldsymbol{u}_t, \boldsymbol{\theta})) \notag \\
& \quad + \frac{1}{2\alpha} \sum_{i=1}^{s_t} \left( h_{t,i}(\boldsymbol{x}_t, \boldsymbol{u}_t, \boldsymbol{\theta}) \right)^2 
) + c_T(\boldsymbol{x}_T, \boldsymbol{\theta}) \notag \\
& \quad + \frac{1}{\alpha} \sum_{i=1}^{q_T} \phi_{\beta}(g_{T,i}(\boldsymbol{x}_T, \boldsymbol{\theta})) 
+ \frac{1}{2\alpha} \sum_{i=1}^{s_T} \left( h_{T,i}(\boldsymbol{x}_T, \boldsymbol{\theta}) \right)^2.
\end{align}
}%
Then, we have the following important result about the approximation for $\boldsymbol{\xi}(\boldsymbol{\theta})$ and $\frac{\partial \boldsymbol{\xi}(\boldsymbol{\theta})}{\partial \boldsymbol{\theta}}$ using $\boldsymbol{\Sigma}(\boldsymbol{\theta},\alpha,\beta)$:
\begin{lemma}\label{lemma:approx}
    Let Lemma \ref{lemma:Differentiability} hold. For any small $\alpha,\beta>0$,
    \begin{enumerate}
        \item there exists a local isolated minimizing trajectory $\boldsymbol{\xi}(\boldsymbol{\theta},\alpha,\beta)$ that solves the OC problem \eqref{eq:unconstrain_optiOC}, 
        and satisfies safety constraints in \eqref{eq:optiOC};
        \item $\boldsymbol{\xi}(\boldsymbol{\theta},\alpha,\beta)$ is once-continuous differentiable with respect to $(\boldsymbol{\theta},\alpha,\beta)$, and $\boldsymbol{\xi}(\boldsymbol{\theta},\alpha,\beta)\rightarrow \boldsymbol{\xi}(\boldsymbol{\theta}),\ \textstyle\frac{\partial \boldsymbol{\xi}(\boldsymbol{\theta},\alpha,\beta)}{\partial\boldsymbol{\theta}}\rightarrow \frac{\partial\boldsymbol{\xi}(\boldsymbol{\theta})}{\partial\boldsymbol{\theta}},$
    as $\alpha,\beta\rightarrow 0$.
    \end{enumerate}
\end{lemma}
\noindent The proof of Lemma \ref{lemma:approx} is given in Appendix. The first assertion states that the approximation $\boldsymbol{\xi}(\boldsymbol{\theta},\alpha,\beta)$ is guaranteed to satisfy all constraints with small positive $\alpha,\beta$. The second assertion states that by choosing small positive $\alpha,\beta$, one can use $\boldsymbol{\xi}(\boldsymbol{\theta},\alpha,\beta)$ and $\frac{\partial \boldsymbol{\xi}(\boldsymbol{\theta},\alpha,\beta)}{\partial\boldsymbol{\theta}}$ of $\boldsymbol{\Sigma}(\boldsymbol{\theta},\alpha,\beta)$ to approximate $\boldsymbol{\xi}(\boldsymbol{\theta})$ and $\frac{\partial\boldsymbol{\xi}(\boldsymbol{\theta})}{\partial\boldsymbol{\theta}}$ of the original constrained system $\boldsymbol{\Sigma}(\boldsymbol{\theta})$, respectively.
Define the Hamiltonian for  $\boldsymbol{\Sigma}(\boldsymbol{\theta},\alpha,\beta)$ as
{\small
\begin{align}\label{eq:Hamiltonian}
\begin{split}
\bar{\boldsymbol{\mathcal{L}}}_t = & c_t(\boldsymbol{x}_t,\boldsymbol{u}_t,\boldsymbol{\theta}) + \boldsymbol{\lambda}_{t+1}'\boldsymbol{f}(\boldsymbol{x}_t,\boldsymbol{u}_t,\boldsymbol{\theta}) + \\
& \frac{1}{\alpha}\sum_{i=1}^{q_t}\phi_{\beta}(g_{t,i}(\boldsymbol{x}_t,\boldsymbol{u}_t,\boldsymbol{\theta})) + \frac{1}{2\alpha}\sum_{i=1}^{s_t}h_{t,i}(\boldsymbol{x}_t,\boldsymbol{u}_t,\boldsymbol{\theta})^2,\\
\bar{\boldsymbol{\mathcal{L}}}_T = & c_T(\boldsymbol{x}_T,\boldsymbol{\theta}) + \frac{1}{\alpha}\textstyle\sum_{i=1}^{q_T}\phi_{\beta}(g_{T,i}(\boldsymbol{x}_T,\boldsymbol{\theta})) + \\
&\frac{1}{2\alpha}\textstyle\sum_{i=1}^{s_T}(h_{T,i}(\boldsymbol{x}_T,\boldsymbol{\theta}))^2.
\end{split}
\end{align}
}%
One can then directly apply PDP in the following Lemma to compute the gradient $\frac{\partial \boldsymbol{\xi}(\boldsymbol{\theta},\alpha,\beta)}{\partial\boldsymbol{\theta}}$.

\begin{lemma}\label{lemma:PDP}
\textbf{(PDP \cite{jin2020pontryagin}).}  
Let $\boldsymbol{X}_t=\tfrac{\partial\boldsymbol{x}_t}{\partial\boldsymbol{\theta}}$, 
$\boldsymbol{U}_t=\tfrac{\partial\boldsymbol{u}_t}{\partial\boldsymbol{\theta}}$, and assume $\bar{\boldsymbol{\mathcal{L}}}_t^{uu}$ is invertible for $t=0,\dots,T-1$.  
Define the recursions
{\small
\begin{align*}
\begin{aligned}
\boldsymbol{V}_t &= \boldsymbol{B}_t + \boldsymbol{A}_t'(\boldsymbol{I}+\boldsymbol{V}_{t+1}\boldsymbol{C}_t)^{-1}\boldsymbol{V}_{t+1}\boldsymbol{A}_t,\\
\boldsymbol{W}_t &= \boldsymbol{A}_t'(\boldsymbol{I}+\boldsymbol{V}_{t+1}\boldsymbol{C}_t)^{-1}(\boldsymbol{W}_{t+1}+\boldsymbol{V}_{t+1}\boldsymbol{M}_t)+\boldsymbol{N}_t,
\end{aligned}
\end{align*}
}%
with $\boldsymbol{V}_T=\bar{\boldsymbol{\mathcal{L}}}_T^{xx}$, $\boldsymbol{W}_T=\bar{\boldsymbol{\mathcal{L}}}_T^{x\theta}$.  
Here, 
{\small
\[
\begin{aligned}
\boldsymbol{A}_t &= \boldsymbol{F}_t^x-\boldsymbol{F}_t^u(\bar{\boldsymbol{\mathcal{L}}}_t^{uu})^{-1}\bar{\boldsymbol{\mathcal{L}}}_t^{ux}, \quad
\boldsymbol{B}_t = \bar{\boldsymbol{\mathcal{L}}}_t^{xx}-\bar{\boldsymbol{\mathcal{L}}}_t^{xu}(\bar{\boldsymbol{\mathcal{L}}}_t^{uu})^{-1}\bar{\boldsymbol{\mathcal{L}}}_t^{ux}, \\
\boldsymbol{C}_t &= \boldsymbol{F}_t^u(\bar{\boldsymbol{\mathcal{L}}}_t^{uu})^{-1}\boldsymbol{F}_t^{u\prime}, \quad
\boldsymbol{M}_t = \boldsymbol{F}_t^{\theta}-\boldsymbol{F}_t^u(\bar{\boldsymbol{\mathcal{L}}}_t^{uu})^{-1}\bar{\boldsymbol{\mathcal{L}}}_t^{u\theta}, \\
\boldsymbol{N}_t &= \bar{\boldsymbol{\mathcal{L}}}_t^{x\theta}-\bar{\boldsymbol{\mathcal{L}}}_t^{xu}(\bar{\boldsymbol{\mathcal{L}}}_t^{uu})^{-1}\bar{\boldsymbol{\mathcal{L}}}_t^{u\theta}.
\end{aligned}
\]
}%
$\bar{\boldsymbol{\mathcal{L}}}_t^{xx}$ denotes the second-order $\bar{\boldsymbol{\mathcal{L}}}_t$ with respect to $\boldsymbol{x}$, with similar notation for other derivatives. $\boldsymbol{F}_t^x,\boldsymbol{F}_t^u,$ and $\boldsymbol{F}_t^{\theta}$ denote first-order derivatives of the dynamics $\boldsymbol{f}$ in \eqref{eq:optiOC}. Then $\tfrac{\partial\boldsymbol{\xi}}{\partial\boldsymbol{\theta}}=\{\boldsymbol{X}_{0:T},\boldsymbol{U}_{0:T-1}\}$ is obtained by iterating ($\boldsymbol{X}_0=\boldsymbol{0}$):
{\small
\begin{align*}
\begin{aligned}
\boldsymbol{U}_t &= -(\bar{\boldsymbol{\mathcal{L}}}_t^{uu})^{-1}\Big(\bar{\boldsymbol{\mathcal{L}}}_t^{ux}\boldsymbol{X}_t+\bar{\boldsymbol{\mathcal{L}}}_t^{u\theta}+\boldsymbol{F}_t^{u\prime}(\boldsymbol{I}+\boldsymbol{V}_{t+1}\boldsymbol{C}_t)^{-1}\\
&\quad(\boldsymbol{V}_{t+1}\boldsymbol{A}_t\boldsymbol{X}_t+\boldsymbol{V}_{t+1}\boldsymbol{M}_t+\boldsymbol{W}_{t+1})\Big),\\
\boldsymbol{X}_{t+1} &= \boldsymbol{F}_t^x \boldsymbol{X}_t + \boldsymbol{F}_t^u \boldsymbol{U}_t + \boldsymbol{F}_t^{\theta}.
\end{aligned}
\end{align*}
}%
\end{lemma}

\subsection{Safe OCIL Algorithm}


With the online parameter estimator and the gradient approximator, the Safe OCIL is summarized as Alg. \ref{algorithm:Safe-OCIL}.

\begin{algorithm}
\caption{Safe Online Control-Informed Learning} \label{algorithm:Safe-OCIL}
\textbf{Set: } $\hat{\boldsymbol{\theta}}_0$, $\boldsymbol{P}_0$, $\alpha$, $\beta$. Convert $\boldsymbol{\Sigma}(\hat{\boldsymbol{\theta}})$ to $\boldsymbol{\Sigma}(\hat{\boldsymbol{\theta}},\alpha,\beta)$.\\
\For{$t=1,..,T$}
{
Obtain new measurement $\boldsymbol{y}_t^*$.\\
Solve $\boldsymbol{\xi}(\hat{\boldsymbol{\theta}}_{t-1})$ w/ $\boldsymbol{\Sigma}(\hat{\boldsymbol{\theta}}_{t-1},\alpha,\beta)$.\\
Obtain $\frac{\partial\boldsymbol{\xi}_t(\hat{\boldsymbol{\theta}}_{t-1})}{\partial\hat{\boldsymbol{\theta}}_{t-1}}$ by Lemma \ref{lemma:PDP}. Obtain $\boldsymbol{L}_t$ via \eqref{eq:chain}.\\
Update $\hat{\boldsymbol{\theta}}_t$ using \eqref{eq:EKF}.\\ 
}
\end{algorithm}



\begin{theorem}\label{theorem}
Suppose the conditions in Lemma~\ref{lemma:Differentiability} hold. 
For sufficiently small $\alpha, \beta > 0$, Lemma~\ref{lemma:approx} 
follows when Lemma~\ref{lemma:PDP} is applied to compute the gradient of 
$\boldsymbol{\Sigma}(\hat{\boldsymbol{\theta}}_t,\alpha,\beta)$. 
Consequently, under the conditions of Lemma~\ref{lemma:OCIL}, 
Algorithm~\ref{algorithm:Safe-OCIL} ensures that the parameter estimate 
$\hat{\boldsymbol{\theta}}_t$ converges locally to $\boldsymbol{\theta}^*$, with $\lim_{t \to \infty} L(\boldsymbol{\xi}(\hat{\boldsymbol{\theta}}_t)) \rightarrow 0$
while guaranteeing that the safety constraints remain satisfied.
\end{theorem}

Assume the conditions in Lemma~\ref{lemma:Differentiability} hold. For small $\alpha,\beta>0$, Lemma~\ref{lemma:approx} ensures that the trajectory $\boldsymbol{\xi}(\boldsymbol{\theta},\alpha,\beta)$ and its gradient $\tfrac{\partial \boldsymbol{\xi}(\boldsymbol{\theta},\alpha,\beta)}{\partial\boldsymbol{\theta}}$ of the unconstrained system $\boldsymbol{\Sigma}(\boldsymbol{\theta},\alpha,\beta)$, computed via PDP in Lemma~\ref{lemma:PDP}, provide accurate approximations of the trajectory $\boldsymbol{\xi}(\boldsymbol{\theta})$ and gradient $\tfrac{\partial \boldsymbol{\xi}(\boldsymbol{\theta})}{\partial\boldsymbol{\theta}}$ of the original constrained system $\boldsymbol{\Sigma}(\boldsymbol{\theta})$. The proof of Lemma~\ref{lemma:approx} is given in Appendix. With these approximations, the online parameter estimator \eqref{eq:EKF} can be applied directly, and convergence is then guaranteed under the conditions of Lemma~\ref{lemma:OCIL}.

\section{Numerical Experiments}

\begin{figure}
\centering
\subfloat[Cartpole Trajectory]
{\label{fig:cp-trajectory}
\includegraphics[width=0.49\linewidth]{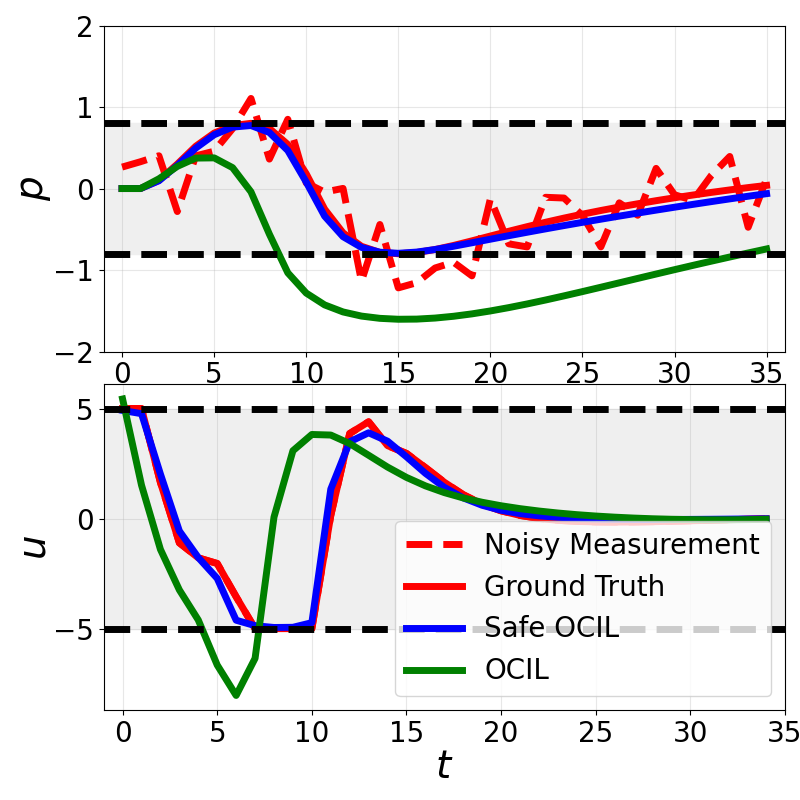}}%
\subfloat[Robot-arm Trajectory]
{\label{fig:ra-trajectory}
\includegraphics[width=0.49\linewidth]{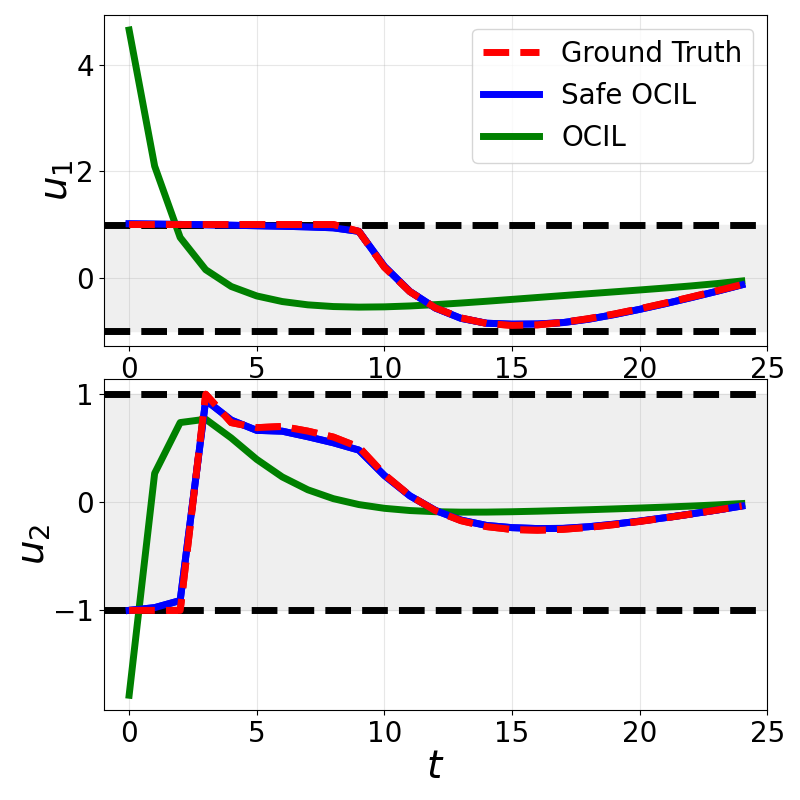}}%
\caption{Trajectories for OCIL and Safe OCIL.}
\label{fig:simulation}
\end{figure} 
\vspace{-0.5em}
\begin{figure}
\centering
{\label{fig:loss}
\includegraphics[width=0.98\linewidth]{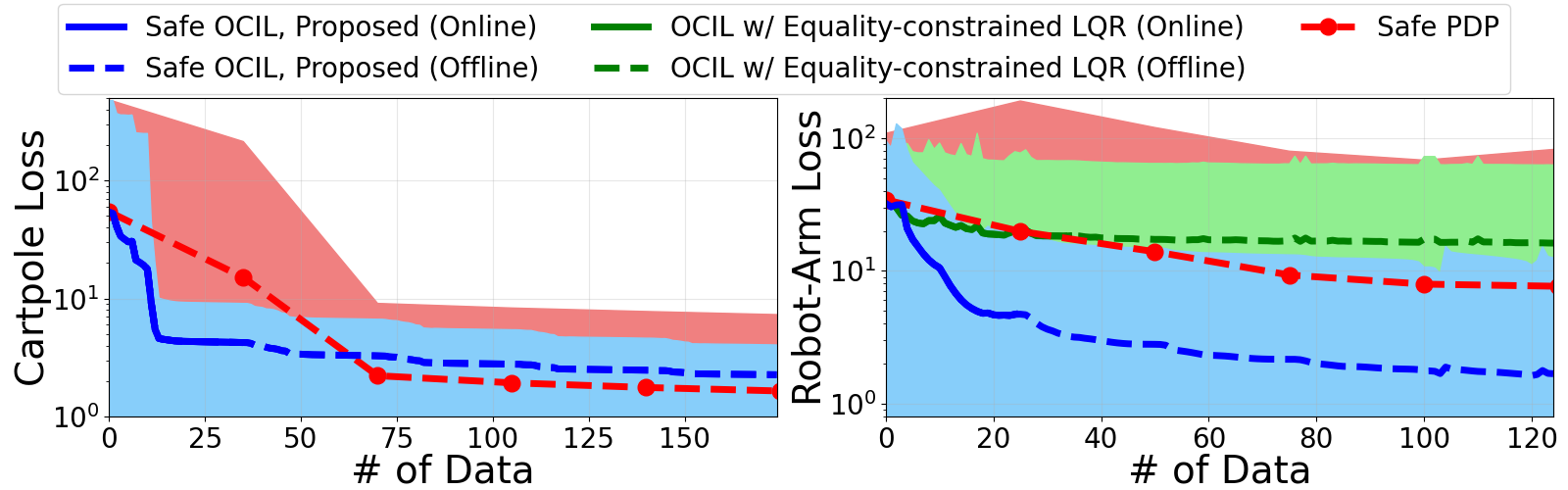}}%
\caption{Loss comparison. Red dots mark update points for Safe PDP, and shaded buffers show loss variation within three standard deviations. The green line represents the results when the Equality-constrained LQR technique is applied to OCIL.}
\label{fig:Loss}
\end{figure} 



Numerical experiments are conducted for a cartpole and a two-link robot-arm system.
The robot-arm system can be found in \cite{spong2008robot}, p. 171, with state vector $\boldsymbol{x}=[q_1,\dot{q}_1,q_2,\dot{q}_2]'$ with $q_1,q_2\in\mathbb{R}$ the joint angles and $\dot{q}_1,\dot{q}_2\in\mathbb{R}$ the joint angular velocities, and the control input $\boldsymbol{u}\in\mathbb{R}^2$ is the vector of torques applied to each joint. The dynamical parameters are $\boldsymbol{\theta}_{\text{dyn}}=[m_1,m_2,l_1,l_2]'$, where $m_1,m_2\in\mathbb{R}$ and $l_1,l_2\in\mathbb{R}$ are the mass and length of each link, respectively. The control costs are $c_t = 0.01||\boldsymbol{u}_t||_2^2 + ||\boldsymbol{\theta}_{\text{obj}}'(\boldsymbol{x}_t-\boldsymbol{x}_{\text{goal}})||_2^2$ and $c_T = ||\boldsymbol{\theta}_{\text{obj}}'(\boldsymbol{x}_T-\boldsymbol{x}_{\text{goal}})||_2^2$, where $\boldsymbol{x}_{\text{goal}}$ is the goal state. The constraint parameters are $\boldsymbol{\theta}_{\text{cstr}}=[u_{\text{max}},q_{\text{max}}]'$, where $u_{\text{max}}\in\mathbb{R}$ is the control limit and $q_{\text{max}}\in\mathbb{R}$ is the joint angle limit. The tunable parameters is given as $\boldsymbol{\theta}=[\boldsymbol{\theta}_{\text{dyn}}',\boldsymbol{\theta}_{\text{obj}}',\boldsymbol{\theta}_{\text{cstr}}']'$. 
The cartpole system from \cite{liang2025online} uses a control cost similar to the robot arm and is constrained by cart position and control input limits. 
For comparison with the offline method, the demonstration is fed repeatedly to Safe OCIL so that once $T$ is reached, the observation at time~0 is given again. 

Fig.~\ref{fig:simulation} compares the control and state trajectories, showing that OCIL fails to enforce safety constraints, whereas Safe OCIL guarantees them and its trajectory aligns well with the ground truth $\boldsymbol{\xi}^*$.
The measurement noise follows a Gaussian distribution \(\mathcal{N}(0,\sigma^2)\) with \(\sigma=0.3\), a substantial uncertainty relative to the \(\pm0.8\,\text{m}\) constraint limit, highlighting the challenge posed by high noise levels. For the cartpole system, $\alpha=0.3$, $\beta=0.075$; for the robot-arm system, $\alpha=0.08$, $\beta=0.02$. 
Fig.~\ref{fig:Loss} reports loss comparisons between Safe OCIL and Safe PDP over 100 random tests. Safe OCIL outperforms Safe PDP in both systems by achieving faster convergence. 
For the robot-arm system, we compare the results obtained when the gradient is computed via the Equality-constrained LQR technique \cite{jin2021safe}. The Equality-constrained LQR–based approach exhibits numerical instability due to errors in identifying active inequality constraints and discontinuities in the trajectory caused by switching active constraints. The proposed gradient approximator addresses this issue by incorporating the constraints into the objective function, thereby eliminating the need to identify active constraints.
In this comparison, $\alpha=\beta=0.1$ is used.


\begin{table}
    \centering
    \footnotesize
    \setlength{\tabcolsep}{3pt}
    \caption{Constraint violations of control input $u$ and cart position $p$.}
    \begin{tabular}{|c|c|>{\centering\arraybackslash}m{1cm}|>{\centering\arraybackslash}m{1cm}|
                        >{\centering\arraybackslash}m{1cm}|>{\centering\arraybackslash}m{1cm}|}
    \hline
    \multirow{2}{*}{\makecell{Noise \\ Level}} & \multirow{2}{*}{Method} 
    & \multicolumn{2}{c|}{\% of Violation} 
    & \multicolumn{2}{c|}{Max. Violation} \\
    \cline{3-6}
    & & $u$ & $p$ & $u$ & $p$ \\
    \hline
    \multirow{2}{*}{0.0} & \multirow{2}{*}{\makecell{\textbf{Safe OCIL} \\ OCIL}} & \textbf{0} & \textbf{0} & \textbf{0} & \textbf{0} \\
    \cline{2-6}
     & & 14.3 & 77.1 & 81.0 & 135.9 \\
     \hline
    \multirow{2}{*}{0.3} & \multirow{2}{*}{\makecell{\textbf{Safe OCIL} \\ OCIL}} & \textbf{0} & \textbf{0} & \textbf{0} & \textbf{0} \\
    \cline{2-6}
     & & 11.4 & 71.4 & 60.8 & 100.4 \\
     \hline
     \multirow{2}{*}{0.6} & \multirow{2}{*}{\makecell{\textbf{Safe OCIL} \\ OCIL}} & \textbf{0} & \textbf{0} & \textbf{0} & \textbf{0} \\
    \cline{2-6}
     & & 11.4 & 77.1 & 63.1 & 107.5 \\
     \hline
    \end{tabular}
    \label{tab:violation}
\end{table}

\begin{table}
    \centering
    \footnotesize
    \setlength{\tabcolsep}{3pt}
    \caption{Computation time [ms].}
    \begin{tabular}{|c|c|>{\centering\arraybackslash}m{2cm}|>{\centering\arraybackslash}m{2cm}|}
    \hline
    System & Method 
    & Per iteration 
    & Gradient \\
    \hline
    \multirow{2}{*}{Cartpole} & \multirow{2}{*}{\makecell{\textbf{Safe OCIL} \\ OCIL}} & $23.55\pm15.24$ & $6.06\pm0.32$ \\
    \cline{2-4}
     & & $22.36\pm5.48$ & $6.30\pm0.61$ \\
     \hline
    \multirow{2}{*}{Robot-arm} & \multirow{2}{*}{\makecell{\textbf{Safe OCIL} \\ OCIL}} & $17.96\pm3.68$ & $5.14\pm0.55$ \\
    \cline{2-4}
     & & $15.76\pm3.86$ & $5.04\pm0.56$ \\
     \hline
    \end{tabular}
    \label{tab:time}
\end{table}

Table \ref{tab:violation} compares constraint violations of cartpole between Safe OCIL and OCIL under different levels of Gaussian noise. "\% of Violation" indicates the portion of the trajectory that violates the constraints, while "Max. Violation" represents the maximum percentage by which the constraints are exceeded. 
This demonstrates that Safe OCIL maintains constraint satisfaction reasonably well, even under high uncertainty. 
Table \ref{tab:time} presents the computation time per iteration and for gradient approximation. 
All results are averaged over $100$ trials on an i9-14900K CPU with a Python implementation using CasADi and the IPOPT solver. 
To be considered online, parameter updates are performed during trajectory execution within the observation update interval, rather than at every control cycle. Specifically, observations are received every $100$~ms for the cartpole system and every $200$~ms for the robot-arm system. The computation times fall within these intervals, confirming the online capability of Safe OCIL.

The proposed algorithm has the following limitations, which motivate further research in the future:
1) $\alpha$ and $\beta$ control the smoothness of the approximation. 
Smaller values enforce the constraints more strictly and make the unconstrained formulation closely match the original constrained problem, but they also increase the curvature (i.e., Hessian magnitude) near active constraints and may cause numerical instability in the solver. 
In practice, $\alpha$ and $\beta$ should be selected empirically to balance accuracy and stability. 
2) Since Safe-OCIL is based on first-order gradients, it can only guarantee convergence to local minima for non-convex optimal control problems.

\section{Conclusion}

This paper presented Safe Online Control-Informed Learning for safety-critical autonomous systems. Using an online 
parameter estimator and gradient approximation, Safe OCIL enables real-time learning under uncertainty while maintaining 
constraint satisfaction. Theoretical analysis established convergence, safety, and gradient validity, and experiments on two systems confirmed its practicality and robustness. 


\bibliographystyle{IEEEtran}
\bibliography{reference}
\appendix
\renewcommand{\thesection}{S}
\renewcommand{\theequation}{\thesection.\arabic{equation}}
\renewcommand{\thelemma}{\thesection.\arabic{lemma}}
\setcounter{equation}{0}
\setcounter{lemma}{0}
\section{Appendix}

Lemma \ref{lemma:approx} is proved by showing the Constrained Pontryagin Maximum/Minimum Principle (C-PMP) conditions and the second-order condition
{\footnotesize
\begin{equation}\label{eq:second_order}
    \sum_{t=0}^{T-1}
    \begin{bmatrix}
        \boldsymbol{x}_t\\ \boldsymbol{u}_t
    \end{bmatrix}'
    \begin{bmatrix}
        \boldsymbol{\mathcal{L}}_t^{xx} & \boldsymbol{\mathcal{L}}_t^{xu} \\
        \boldsymbol{\mathcal{L}}_t^{ux} & \boldsymbol{\mathcal{L}}_t^{uu} \\
    \end{bmatrix}
    \begin{bmatrix}
        \boldsymbol{x}_t\\ \boldsymbol{u}_t
    \end{bmatrix}
    + \boldsymbol{x}_T' \boldsymbol{\mathcal{L}}_{T}^{xx}\boldsymbol{x}_T > 0
\end{equation}
}%
hold for certain non-zero trajectories. The complete C-PMP conditions and trajectory requirements are provided in \cite{jin2021safe}. 
For notational simplicity, $\boldsymbol{\mathcal{L}}_t^{x}$ and $\boldsymbol{\mathcal{L}}_t^{xx}$ denote the first- and second-order derivatives of $\boldsymbol{\mathcal{L}}_t$ with respect to $\boldsymbol{x}$, respectively. Similar notation applies to other Hamiltonian terms.
We first modify the C-PMP condition for $\boldsymbol{\Sigma}(\boldsymbol{\theta},\alpha,\beta)$:
{\footnotesize
\begin{subequations}\label{eq:PMP}
\begin{align}
    & \boldsymbol{x}_{t+1} = \boldsymbol{f}(\boldsymbol{x}_t,\boldsymbol{u}_t,\boldsymbol{\theta}) \quad \text{and} \quad \boldsymbol{x}_0 = \boldsymbol{x}_0(\boldsymbol{\theta}),\label{eq:PMP1}\\
    & \boldsymbol{\lambda}_t = \boldsymbol{\mathcal{L}}_t^{x}(\boldsymbol{x}_t,\boldsymbol{u}_t,\boldsymbol{\lambda}_{t+1},\boldsymbol{\mu}_t,\boldsymbol{\nu}_t,\boldsymbol{\theta}), \ \boldsymbol{\lambda}_T = \boldsymbol{\mathcal{L}}_{T}^{x}(\boldsymbol{x}_T,\boldsymbol{\mu}_T,\boldsymbol{\nu}_T,\boldsymbol{\theta}),\label{eq:PMP2}\\
    & \boldsymbol{0} = \boldsymbol{\mathcal{L}}_{t}^{u}(\boldsymbol{x}_t,\boldsymbol{u}_t,\boldsymbol{\lambda}_{t+1},\boldsymbol{\mu}_t,\boldsymbol{\nu}_t,\boldsymbol{\theta}),\label{eq:PMP4}\\
    & h_{t,i}(\boldsymbol{x}_t,\boldsymbol{u}_t,\boldsymbol{\theta}) = \nu_{t,i} \alpha, \quad i=1,2,...,s_t,\label{eq:PMP5}\\
    & h_{T,i}(\boldsymbol{x}_T,\boldsymbol{\theta}) = \nu_{T,i} \alpha, \quad i=1,2,...,s_T,\label{eq:PMP6}\\
    & \mu_{t,i} g_{t,i}(\boldsymbol{x}_t,\boldsymbol{u}_t,\boldsymbol{\theta}) = 0, \quad i=1,2,...,q_t,\label{eq:PMP7}\\
    & \mu_{T,i} g_{T,i}(\boldsymbol{x}_T,\boldsymbol{\theta}) = 0, \quad i=1,2,...,q_T.\label{eq:PMP8}
\end{align}
\end{subequations}
}%
where \ref{eq:PMP1}-\ref{eq:PMP4} are the same with C-PMP conditions. Let
{\footnotesize
\begin{subequations}\label{eq:mu}
\begin{align}
    & \mu_{t,i} = \frac{1}{\alpha}\phi_{\beta}'(g_{t,i}(\boldsymbol{x}_t,\boldsymbol{u}_t,\boldsymbol{\theta})) \geq 0, \quad i=1,2,...,q_t,\label{eq:mu_t}\\
    & \mu_{T,i} = \frac{1}{\alpha}\phi_{\beta}'(g_{T,i}(\boldsymbol{x}_T,\boldsymbol{\theta})) \geq 0, \quad i=1,2,...,q_T.\label{eq:mu_T}
\end{align}
\end{subequations}
}%
For $\boldsymbol{\theta}$ near $\boldsymbol{\theta}^*$ and $\beta \rightarrow 0$, the multiplier
{\footnotesize
\begin{equation}
    \mu_{t,i} = \frac{1}{\alpha}\frac{1}{1+e^{-g_{t,i}(\boldsymbol{x}_t,\boldsymbol{u}_t,\boldsymbol{\theta})/\beta}}\rightarrow 0
\end{equation}
}%
if an inequality constraint is inactive; and
{\footnotesize
\begin{equation}
    \mu_{t,i} = \frac{1}{\alpha}\frac{1}{1+e^{0/\beta}} \rightarrow \frac{1}{2\alpha}>0
\end{equation}
}%
if an inequality constraint is active.
Similar procedures apply to \eqref{eq:PMP8}. 
By letting $\alpha,\beta \rightarrow 0$ and $\boldsymbol{\theta} = \bar{\boldsymbol{\theta}}$, \eqref{eq:PMP5}-\eqref{eq:PMP8} coincide with the C-PMP condition for $\boldsymbol{\Sigma}(\bar{\boldsymbol{\theta}})$.

We compactly express the modified C-PMP conditions as $\boldsymbol{\mathcal{M}}(\boldsymbol{\xi},\boldsymbol{\lambda},\boldsymbol{\mu},\boldsymbol{\nu};
      \boldsymbol{\theta},\alpha,\beta)=\boldsymbol{0}$,
where $(\boldsymbol{\xi},\boldsymbol{\lambda},\boldsymbol{\mu},\boldsymbol{\nu})$ are the unknowns and 
$(\boldsymbol{\theta},\alpha,\beta)$ are parameters.  
At $(\bar{\boldsymbol{\theta}},0,0)$, Lemma~\ref{lemma:Differentiability} guarantees a solution 
$(\bar{\boldsymbol{\xi}},\bar{\boldsymbol{\lambda}},\bar{\boldsymbol{\mu}},\bar{\boldsymbol{\nu}})$ 
to the original constrained system $\boldsymbol{\Sigma}(\bar{\boldsymbol{\theta}})$, i.e., $\boldsymbol{\mathcal{M}}(\bar{\boldsymbol{\xi}},\bar{\boldsymbol{\lambda}},\bar{\boldsymbol{\mu}},\bar{\boldsymbol{\nu}},
\bar{\boldsymbol{\theta}},0,0)=\boldsymbol{0}$.
Because the dynamics, cost, and constraints are \(C^2\), 
$\boldsymbol{\mathcal{M}}$ is continuously differentiable in both variables and parameters.  
Lemma~\ref{lemma:Differentiability} ensures that the Jacobian of 
$\boldsymbol{\mathcal{M}}$ with respect to $(\boldsymbol{\xi},\boldsymbol{\lambda},\boldsymbol{\mu},\boldsymbol{\nu})$ is nonsingular.  
For notation simplicity, we denote by $\tilde{\cdot}$ for the solution of $\boldsymbol{\Sigma}(\boldsymbol{\theta},\alpha,\beta)$, 
e.g., $\tilde{\boldsymbol{x}}_t$ for $\boldsymbol{x}_t(\boldsymbol{\theta},\alpha,\beta)$.  
Applying the \emph{Implicit Function Theorem} \cite{rudin1976principles} to \eqref{eq:PMP} Lemma \ref{lemma:diff_of_xi}:
\begin{alemma}\label{lemma:diff_of_xi}
    For any ($\boldsymbol{\theta},\alpha,\beta$) within a neighborhood of ($\bar{\boldsymbol{\theta}},0,0$), there exists a unique \(C^1\) function $(\tilde{\boldsymbol{\xi}}, \tilde{\boldsymbol{\lambda}}_{1:T},\tilde{\boldsymbol{\mu}}_{0:T},\tilde{\boldsymbol{\nu}}_{0:T})$ which satisfies \eqref{eq:PMP} and coincide with function ($\boldsymbol{\xi}(\bar{\boldsymbol{\theta}}), \boldsymbol{\lambda}_{1:T}(\bar{\boldsymbol{\theta}}),\boldsymbol{\mu}_{0:T}(\bar{\boldsymbol{\theta}}),\boldsymbol{\nu}_{0:T}(\bar{\boldsymbol{\theta}})$) at $ (\bar{\boldsymbol{\theta}},0,0)$.
\end{alemma}

\noindent\textbf{Proof of the second part of assertion 1.}
We need to show that $\tilde{\boldsymbol{\xi}}$ satisfies the safety constraints in \eqref{eq:optiOC}.
For inactive constraints $g_{t,i}(\boldsymbol{x}_t(\bar{\boldsymbol{\theta}}), \boldsymbol{u}_t(\bar{\boldsymbol{\theta}}), \bar{\boldsymbol{\theta}})<0$, due to the continuity of $g_{t,i}$ and $\tilde{\boldsymbol{\xi}}$, $g_{t,i}(\tilde{\boldsymbol{x}}_t, \tilde{\boldsymbol{u}}_t, \boldsymbol{\theta}) \rightarrow g_{t,i}(\boldsymbol{x}_t(\bar{\boldsymbol{\theta}}), \boldsymbol{u}_t(\bar{\boldsymbol{\theta}}), \bar{\boldsymbol{\theta}})<0$,
as $(\boldsymbol{\theta},\alpha,\beta) \rightarrow (\bar{\boldsymbol{\theta}},0,0)$. For active constraints $g_{t,i}(\boldsymbol{x}_t(\bar{\boldsymbol{\theta}}), \boldsymbol{u}_t(\bar{\boldsymbol{\theta}}), \bar{\boldsymbol{\theta}})=0$ and $\mu_{t,i}(\bar{\boldsymbol{\theta}})>0$ (strict complementarity), due to the continuity of $\tilde{\mu}_{t,i}$, $\tilde{\mu}_{t,i} \rightarrow \mu_{t,i}(\bar{\boldsymbol{\theta}}) > 0$ as $(\boldsymbol{\theta},\alpha,\beta) \rightarrow (\bar{\boldsymbol{\theta}},0,0)$.
From \eqref{eq:mu_t}, we obtain $g_{t,i}(\boldsymbol{x}_t,\boldsymbol{u}_t,\boldsymbol{\theta}) = \beta \ln (\frac{\alpha \mu_{t,i}}{1 - \alpha \mu_{t,i}})$.
Then, the sufficient condition for $g_{t,i}(\boldsymbol{x}_t,\boldsymbol{u}_t,\boldsymbol{\theta}) < 0$ is $\alpha \mu_{t,i} < \frac{1}{2}$. Thus, $g_{t,i}(\tilde{\boldsymbol{x}}_t, \tilde{\boldsymbol{u}}_t, \boldsymbol{\theta}) < 0$ for any $(\boldsymbol{\theta},\alpha,\beta)$ near $(\bar{\boldsymbol{\theta}},0,0)$. By a similar proof for other constraints, we conclude that $\tilde{\boldsymbol{\xi}}$ satisfies the safety constraints for any $(\boldsymbol{\theta},\alpha,\beta)$ near $(\bar{\boldsymbol{\theta}},0,0)$ 
with small $\alpha,\beta>0$.

\noindent\textbf{Proof of the first part of assertion 1.}
We will show for any $(\boldsymbol{\theta},\alpha,\beta)$ near $(\bar{\boldsymbol{\theta}},0,0)$ with $\alpha>0$ and $\beta>0$, $\tilde{\boldsymbol{\xi}}$ is a local isolated minimizing trajectory to $\boldsymbol{\Sigma}(\boldsymbol{\theta},\alpha,\beta)$. By plugging \eqref{eq:PMP5} and \eqref{eq:mu_t} into the \eqref{eq:PMP2}, 
we have 

{\footnotesize
\begin{align*}
\begin{split}
    \bar{\boldsymbol{\mathcal{L}}}_t^{x}(\tilde{\boldsymbol{x}}_t,\tilde{\boldsymbol{u}}_t,\boldsymbol{\theta})
    = & \nabla c_t(\tilde{\boldsymbol{x}}_t,\tilde{\boldsymbol{u}}_t,\boldsymbol{\theta}) + \nabla\tilde{\boldsymbol{\lambda}}_{t+1}'\boldsymbol{f}(\tilde{\boldsymbol{x}}_t,\tilde{\boldsymbol{u}}_t,\boldsymbol{\theta})\\
    &  + \sum_{i=1}^{q_t} \frac{1}{\alpha}\phi_{\beta}'(g_{t,i}(\tilde{\boldsymbol{x}}_t,\tilde{\boldsymbol{u}}_t,\boldsymbol{\theta}))\nabla\phi_{\beta}(g_{t,i}(\tilde{\boldsymbol{x}}_t,\tilde{\boldsymbol{u}}_t,\boldsymbol{\theta}))\\
    & + \sum_{i=1}^{s_t}\frac{1}{\alpha}h_{t,i}(\tilde{\boldsymbol{x}}_t,\tilde{\boldsymbol{u}}_t,\boldsymbol{\theta})\nabla h_{t,i}(\tilde{\boldsymbol{x}}_t,\tilde{\boldsymbol{u}}_t,\boldsymbol{\theta}) = \tilde{\boldsymbol{\lambda}}_t .
\end{split}
\end{align*}
}%
With similar procedures, 
$\tilde{\boldsymbol{x}}_{t+1} = \boldsymbol{f}(\tilde{\boldsymbol{x}}_t,\tilde{\boldsymbol{u}}_t,\boldsymbol{\theta})$, $\tilde{\boldsymbol{x}}_0 = \boldsymbol{x}_0(\boldsymbol{\theta})$, $\tilde{\boldsymbol{\lambda}}_t = \bar{\boldsymbol{\mathcal{L}}}_t^{x}(\tilde{\boldsymbol{x}}_t,\tilde{\boldsymbol{u}}_t,\tilde{\boldsymbol{\lambda}}_{t+1},(\boldsymbol{\theta},\alpha,\beta))$, $\tilde{\boldsymbol{\lambda}}_T = \bar{\boldsymbol{\mathcal{L}}}_T^{x}(\tilde{\boldsymbol{x}}_T,(\boldsymbol{\theta},\alpha,\beta))$, $\boldsymbol{0} = \bar{\boldsymbol{\mathcal{L}}}_t^{u}(\tilde{\boldsymbol{x}}_t,\tilde{\boldsymbol{u}}_t,\tilde{\boldsymbol{\lambda}}_{t+1},(\boldsymbol{\theta},\alpha,\beta)).$
This is exactly the PMP conditions for $\boldsymbol{\Sigma}(\boldsymbol{\theta},\alpha,\beta)$ with its Hamiltonian \eqref{eq:Hamiltonian}, indicating that $\tilde{\boldsymbol{\xi}}$ satisfies the PMP condition for $\boldsymbol{\Sigma}(\boldsymbol{\theta},\alpha,\beta)$. To show $\tilde{\boldsymbol{\xi}}$ is a local isolated minimizing trajectory to $\boldsymbol{\Sigma}(\boldsymbol{\theta},\alpha,\beta)$ for any $(\boldsymbol{\theta},\alpha,\beta)$ near $(\bar{\boldsymbol{\theta}},0,0)$ with $\alpha,\beta > 0$, we only need to show it satisfies the second-order condition in \eqref{eq:second_order}.
The second-order derivatives of $\bar{\boldsymbol{\mathcal{L}}}_t$ and $\bar{\boldsymbol{\mathcal{L}}}_T$ are
{\footnotesize
\begin{align*}
\begin{split}
    \nabla^2\bar{\boldsymbol{\mathcal{L}}}_t=& 
    \nabla^2\boldsymbol{\mathcal{L}}_t
     + \sum_{i=1}^{q_t}\frac{1}{\alpha\beta}\frac{e^{g_{t,i}/\beta}}{(1+e^{g_{t,i}/\beta})^2} \nabla^2 g_{t,i} + \sum_{i=1}^{s_t}\frac{1}{\alpha}\nabla^2 h_{t,i},\\
    \bar{\boldsymbol{\mathcal{L}}}_T^{xx} = & \boldsymbol{\mathcal{L}}_T^{xx} + \sum_{i=1}^{q_T}\frac{1}{\alpha\beta}\frac{e^{g_{t,i}/\beta}}{(1+e^{g_{t,i}/\beta})^2}\nabla^2g_{T,i} + \sum_{i=1}^{s_T}\frac{1}{\alpha}\nabla^2h_{T,i}.
\end{split}
\end{align*}
}%
Given any $\boldsymbol{x}$ and $\boldsymbol{u}$ with appropriate dimensions, one has
{\footnotesize
\begin{align*}
\begin{split}
&\begin{bmatrix}
    \boldsymbol{x}\\ \boldsymbol{u}
\end{bmatrix}'
\nabla^2\bar{\boldsymbol{\mathcal{L}}}_t
\begin{bmatrix}
    \boldsymbol{x}\\ \boldsymbol{u}
\end{bmatrix} = \begin{bmatrix}
    \boldsymbol{x}\\ \boldsymbol{u}
\end{bmatrix}'
\nabla^2\boldsymbol{\mathcal{L}}_t
\begin{bmatrix}
    \boldsymbol{x}\\ \boldsymbol{u}
\end{bmatrix} + \sum_{i=1}^{q_t}\Big(\frac{1}{\alpha\beta}\frac{e^{g_{t,i}/\beta}}{(1+e^{g_{t,i}/\beta})^2}\\
&(\frac{\partial g_{t,i}}{\partial\boldsymbol{x}_t}\boldsymbol{x}+\frac{\partial g_{t,i}}{\partial\boldsymbol{u}_t}\boldsymbol{u})^2\Big) + \sum_{i=1}^{s_t}\frac{1}{\alpha}(\frac{\partial h_{t,i}}{\partial\boldsymbol{x}_t}\boldsymbol{x}+\frac{\partial h_{t,i}}{\partial\boldsymbol{u}_t}\boldsymbol{u})^2,
\end{split}\\
\begin{split}
&\boldsymbol{x}'\bar{\boldsymbol{\mathcal{L}}}_T^{xx}\boldsymbol{x} = \boldsymbol{x}'\boldsymbol{\mathcal{L}}_T^{xx}\boldsymbol{x} + \sum_{i=1}^{q_T}\frac{1}{\alpha\beta}\frac{e^{g_{T,i}/\beta}}{(1+e^{g_{T,i}/\beta})^2}(\frac{\partial g_{T,i}}{\partial\boldsymbol{x}_T}\boldsymbol{x})^2\\
& + \sum_{i=1}^{s_T}\frac{1}{\alpha}(\frac{\partial h_{T,i}}{\partial\boldsymbol{x}_T}\boldsymbol{x})^2.
\end{split}
\end{align*}
}%
We need to prove:
{\footnotesize
\begin{align}\label{eq:sosc}
\begin{split}
    \sum_{t=0}^{T-1}
    \begin{bmatrix}
        \boldsymbol{x}_t\\ \boldsymbol{u}_t
    \end{bmatrix}'
    \nabla^2\bar{\boldsymbol{\mathcal{L}}}_t(\boldsymbol{\theta},\alpha,\beta)
    \begin{bmatrix}
        \boldsymbol{x}_t\\ \boldsymbol{u}_t
    \end{bmatrix} + \boldsymbol{x}_T'\bar{\boldsymbol{\mathcal{L}}}_T^{xx}(\boldsymbol{\theta},\alpha,\beta)\boldsymbol{x}_T > 0,
\end{split}
\end{align}
}%
for any $\{\boldsymbol{x}_{0:T},\boldsymbol{u}_{0:T-1}\}\neq\boldsymbol{0}$ satisfying
{\footnotesize
\begin{equation}\label{eq:sosc_dyn}
    \boldsymbol{x}_{t+1}=F_t^x (\boldsymbol{\theta},\alpha,\beta) \boldsymbol{x}_t + F_t^u (\boldsymbol{\theta},\alpha,\beta)\boldsymbol{u}_t, \  \boldsymbol{x}_0=\boldsymbol{0}.
\end{equation}
}%
\textbf{Proof by contradiction:} Suppose that the second-order condition in \eqref{eq:sosc}-\eqref{eq:sosc_dyn} is false. There must exist a sequence of parameters $(\boldsymbol{\theta}^{(k)},\alpha^{(k)}>0,\beta^{(k)}>0)$ 
, and a sequence of non-zero trajectory 
such that $(\boldsymbol{\theta}^{(k)},\alpha^{(k)},\beta^{(k)})\rightarrow(\bar{\boldsymbol{\theta}},0,0)$, $\boldsymbol{x}_{t+1}^{(k)}=F_t^x (\boldsymbol{\theta}^{(k)},\alpha^{(k)},\beta^{(k)}) \boldsymbol{x}_t^{(k)} + F_t^u (\boldsymbol{\theta}^{(k)},\alpha^{(k)},\beta^{(k)})\boldsymbol{u}_t^{(k)}$ with $\boldsymbol{x}_0=\boldsymbol{0}$, and 
{\footnotesize
\begin{align}\label{eq:contradiction}
\begin{split}
    &\sum_{t=0}^{T-1}
    \begin{bmatrix}
        \boldsymbol{x}_t^{(k)}\\ \boldsymbol{u}_t^{(k)}
    \end{bmatrix}'
    \nabla^2\bar{\boldsymbol{\mathcal{L}}}_t(\boldsymbol{\theta}^{(k)},\alpha^{(k)},\beta^{(k)})
    \begin{bmatrix}
        \boldsymbol{x}_t^{(k)}\\ \boldsymbol{u}_t^{(k)}
    \end{bmatrix} \\
    &\quad \quad \quad+ \boldsymbol{x}_T^{k\prime}\bar{\boldsymbol{\mathcal{L}}}_T^{xx}(\boldsymbol{\theta}^{(k)},\alpha^{(k)},\beta^{(k)})\boldsymbol{x}_T^{(k)} \leq 0,
\end{split}
\end{align}
}%
for $k=1,2,3...$. Here, $(\boldsymbol{\theta}^{(k)},\alpha^{(k)},\beta^{(k)})$ means these derivatives are evaluated at $\boldsymbol{\xi}(\boldsymbol{\theta}^{(k)},\alpha^{(k)},\beta^{(k)})$.  Without loss of generality, assume $\|\mathrm{col}\{\boldsymbol{x}_{0:T}^{k}, \boldsymbol{u}_{0:T-1}^{k}\}\| = 1$ for all $k$. Select a convergent sub-sequence $\{\boldsymbol{x}_{0:T}^{k}, \boldsymbol{u}_{0:T-1}^{k}\}$ and call its limit $\{\boldsymbol{x}_{0:T}^*, \boldsymbol{u}_{0:T-1}^*\}$, that is, $\{\boldsymbol{x}_{0:T}^{(k)}, \boldsymbol{u}_{0:T-1}^{(k)}\} \to \{\boldsymbol{x}_{0:T}^*, \boldsymbol{u}_{0:T-1}^*\}$, and $(\boldsymbol{\theta}^{(k)}, \alpha^{(k)}, \beta^{(k)}) \to (\bar{\boldsymbol{\theta}}, 0, 0)$ as $k \to +\infty$ and $\boldsymbol{x}_{t+1}^* = F_t^x(\bar{\boldsymbol{\theta}}, 0,0) \boldsymbol{x}_t^* + F_t^u(\bar{\boldsymbol{\theta}}, 0,0)  \boldsymbol{u}_t^*$
with $x_0^* = 0$. 

\noindent \textbf{Case 1:} At least one of the following holds:
{\footnotesize
\begin{align*}
\begin{split}
    &\bar{\boldsymbol{G}}_t^x(\bar{\boldsymbol{\theta}}, 0,0) \boldsymbol{x}_t^* + \bar{\boldsymbol{G}}_t^u(\bar{\boldsymbol{\theta}}, 0,0) \boldsymbol{u}_t^* \neq \boldsymbol{0} \  \exists t \quad \text{or} \quad \bar{\boldsymbol{G}}_T^x(\bar{\boldsymbol{\theta}}, 0,0) \boldsymbol{x}_T^* \neq \boldsymbol{0} \quad \text{or}\\
    &\boldsymbol{H}_t^x(\bar{\boldsymbol{\theta}}, 0,0) \boldsymbol{x}_t^* + \boldsymbol{H}_t^u(\bar{\boldsymbol{\theta}}, 0,0) \boldsymbol{u}_t^* \neq \boldsymbol{0} \  \exists t \quad \text{or} \quad \boldsymbol{H}_T^x(\bar{\boldsymbol{\theta}}, 0,0) \boldsymbol{x}_T^* \neq \boldsymbol{0}.
\end{split}
\end{align*}
}%
$\boldsymbol{H}_t^x$ is the first-order derivative of $\boldsymbol{h}_t$ with respect to $\boldsymbol{x}$ and the similar convention applies to $\boldsymbol{H}_T^x, \boldsymbol{H}_t^u$; $\bar{\boldsymbol{G}}_t^x$, $\bar{\boldsymbol{G}}_t^u$ for $\bar{\boldsymbol{g}}_t$; and $\bar{\boldsymbol{G}}_T^x$ for $\bar{\boldsymbol{g}}_T$,
where
{\footnotesize
\begin{align*}
\begin{split}
&\bar{\boldsymbol{g}}_t(\boldsymbol{x}_t,\boldsymbol{u}_t,\boldsymbol{\theta}) =  \emph{col}\{g_{t,i}|g_{t,i}(\boldsymbol{x}_t(\boldsymbol{\theta}),\boldsymbol{u}_t(\boldsymbol{\theta}),\boldsymbol{\theta})=0, i=1,...,q_t\},\\
&\bar{\boldsymbol{g}}_T(\boldsymbol{x}_T,\boldsymbol{\theta}) =  \emph{col}\{g_{T,i}|g_{T,i}(\boldsymbol{x}_T(\boldsymbol{\theta}),\boldsymbol{\theta})=0, i=1,...,q_T\}.\\
\end{split}
\end{align*}
}%
Here, $\bar{\boldsymbol{g}}_t$ and $\bar{\boldsymbol{g}}_T$ are formed by stacking all active inequality constraints at $\boldsymbol{\xi}(\boldsymbol{\theta})$.
As $k \rightarrow +\infty$, the residual term
{\footnotesize
\begin{align*}\label{eq:case1_inf}
\begin{split}
    &\sum_{t=0}^{T-1}
    \Big(\sum_{i=1}^{q_t}\frac{1}{\alpha^{(k)}\beta^{(k)}}\frac{e^{g_{t,i}^{(k)}/\beta^{(k)}}}{(1+e^{g_{t,i}^{(k)}/\beta^{(k)}})^2}(\frac{\partial g_{t,i}^{(k)}}{\partial\boldsymbol{x}_t}\boldsymbol{x}_t^{(k)} +\frac{\partial g_{t,i}^{(k)}}{\partial\boldsymbol{u}_t}\boldsymbol{u}_t^{(k)})^2\\
    &+ \sum_{i=1}^{s_t}\frac{1}{\alpha^{(k)}}(\frac{\partial h_{t,i}^{(k)}}{\partial\boldsymbol{x}_t}\boldsymbol{x}_t^{(k)}+\frac{\partial h_{t,i}^{(k)}}{\partial\boldsymbol{u}_t}\boldsymbol{u}_t^{(k)})^2\Big) + \sum_{i=1}^{q_T}\Big(\frac{1}{\alpha^{(k)}\beta^{(k)}}\\
    &\frac{e^{g_{T,i}^{(k)}/\beta^{(k)}}}{(1+e^{g_{T,i}^{(k)}/\beta^{(k)}})^2}(\frac{\partial g_{T,i}^{(k)}}{\partial\boldsymbol{x}_T}\boldsymbol{x}_T^{(k)})^2\Big) + \sum_{i=1}^{s_T}\frac{1}{\alpha^{(k)}}(\frac{\partial h_{T,i}^{(k)}}{\partial\boldsymbol{x}_T}\boldsymbol{x}_T^{(k)})^2\rightarrow +\infty,
\end{split}
\end{align*}
}%
where the superscript $k$ denotes the values to be evaluated at $\boldsymbol{\xi}(\boldsymbol{\theta}^{(k)},\alpha^{(k)},\beta^{(k)})$. The above limit is true because at least one of the terms is $+\infty$. Here, we use the fact that for active inequality constraint $g_{t,i}(\boldsymbol{x}_t(\bar{\boldsymbol{\theta}}),\boldsymbol{u}_t(\bar{\boldsymbol{\theta}}),\bar{\boldsymbol{\theta}})=0$, 
{\footnotesize
\begin{equation}
    \frac{1}{\alpha\beta}\frac{e^{g_{t,i}(\tilde{\boldsymbol{x}}_t,\tilde{\boldsymbol{u}}_t,\boldsymbol{\theta})/\beta}}{(1+e^{g_{t,i}(\tilde{\boldsymbol{x}}_t,\tilde{\boldsymbol{u}}_t,\boldsymbol{\theta})/\beta})^2} \rightarrow +\infty
\end{equation}
}%
as $(\boldsymbol{\theta},\alpha,\beta)\rightarrow(\bar{\boldsymbol{\theta}},0,0)$.
This leads the left side of \eqref{eq:contradiction} to be $+\infty$, which contradicts \eqref{eq:contradiction}.

\noindent \textbf{Case 2:} All of the following holds:
{\footnotesize
\begin{align*}
\begin{split}
    &\bar{\boldsymbol{G}}_t^x(\bar{\boldsymbol{\theta}}, 0,0) \boldsymbol{x}_t^* + \bar{\boldsymbol{G}}_t^u(\bar{\boldsymbol{\theta}}, 0,0) \boldsymbol{u}_t^* = \boldsymbol{0} \  \exists t; \quad \bar{\boldsymbol{G}}_T^x(\bar{\boldsymbol{\theta}}, 0,0) \boldsymbol{x}_T^* = \boldsymbol{0};\\
    &\boldsymbol{H}_t^x(\bar{\boldsymbol{\theta}}, 0,0) \boldsymbol{x}_t^* + \boldsymbol{H}_t^u(\bar{\boldsymbol{\theta}}, 0,0) \boldsymbol{u}_t^* = \boldsymbol{0} \  \exists t; \quad \boldsymbol{H}_T^x(\bar{\boldsymbol{\theta}}, 0,0) \boldsymbol{x}_T^* = \boldsymbol{0}.
\end{split}
\end{align*}
}
In this case,
{\footnotesize
\begin{align*}
\begin{split}
    &\lim_{k\rightarrow+\infty}\bigg(\sum_{t=0}^{T-1}
    \begin{bmatrix}
        \boldsymbol{x}_t^{(k)}\\ \boldsymbol{u}_t^{(k)}
    \end{bmatrix}'
    \nabla^2 \bar{\boldsymbol{\mathcal{L}}}_t(\boldsymbol{\theta}^{(k)},\alpha^{(k)},\beta^{(k)})
    \begin{bmatrix}
        \boldsymbol{x}_t^{(k)}\\ \boldsymbol{u}_t^{(k)}
    \end{bmatrix} \\
    & \quad + \boldsymbol{x}_T^{k\prime}\bar{\boldsymbol{\mathcal{L}}}_T^{xx}(\boldsymbol{\theta}^{(k)},\alpha^{(k)},\beta^{(k)})\boldsymbol{x}_T^{(k)}\bigg) \\
    & \quad \geq \lim_{k\rightarrow+\infty}\bigg(\sum_{t=0}^{T-1}
    \begin{bmatrix}
        \boldsymbol{x}_t^{(k)}\\ \boldsymbol{u}_t^{(k)}
    \end{bmatrix}'
    \nabla^2 \boldsymbol{\mathcal{L}}_t(\boldsymbol{\theta}^{(k)},\alpha^{(k)},\beta^{(k)})
    \begin{bmatrix}
        \boldsymbol{x}_t^{(k)}\\ \boldsymbol{u}_t^{(k)}
    \end{bmatrix} \\
    & \quad + \boldsymbol{x}_T^{k\prime}\boldsymbol{\mathcal{L}}_T^{xx}(\boldsymbol{\theta}^{(k)},\alpha^{(k)},\beta^{(k)})\boldsymbol{x}_T^{(k)}\bigg) \\
    & \quad =\sum_{t=0}^{T-1}
    \begin{bmatrix}
        \boldsymbol{x}_t^*\\ \boldsymbol{u}_t^*
    \end{bmatrix}'
    \nabla^2 \boldsymbol{\mathcal{L}}_t(\bar{\boldsymbol{\theta}},0,0)
    \begin{bmatrix}
        \boldsymbol{x}_t^*\\ \boldsymbol{u}_t^*
    \end{bmatrix}  + \boldsymbol{x}_T^{*\prime}\boldsymbol{\mathcal{L}}_T^{xx}(\bar{\boldsymbol{\theta}},0,0)\boldsymbol{x}_T^* > 0,
\end{split}
\end{align*}
}%
because the residual term is always non-negative.
This inequality contradicts \eqref{eq:contradiction}. We can conclude that for any $(\boldsymbol{\theta},\alpha,\beta)$ near $(\bar{\boldsymbol{\theta}},0,0)$ with $\alpha,\beta>0$, the trajectory of the unconstrained optimal control system $\boldsymbol{\Sigma}(\boldsymbol{\theta},\alpha,\beta)$ satisfies its C-PMP condition and the second-order condition. Thus, $\tilde{\boldsymbol{\xi}}$ is a local isolated minimizing trajectory to $\boldsymbol{\Sigma}(\boldsymbol{\theta},\alpha,\beta)$.

\noindent \textbf{Proof of assertion 2.}
Given the conditions of Lemma~\ref{lemma:Differentiability} for $\boldsymbol{\Sigma}(\bar{\boldsymbol{\theta}})$,  
for any $(\boldsymbol{\theta},\alpha,\beta)$ near $(\bar{\boldsymbol{\theta}},0,0)$ there exists a unique \(C^1\) function $(\tilde{\boldsymbol{\xi}},\tilde{\boldsymbol{\lambda}}_{1:T},\tilde{\boldsymbol{\mu}}_{0:T},\tilde{\boldsymbol{\nu}}_{0:T})$ satisfying \eqref{eq:PMP}. 
For small $\alpha,\beta>0$, $\tilde{\boldsymbol{\xi}}$ is a safe local isolated minimizing trajectory of $\boldsymbol{\Sigma}(\boldsymbol{\theta},\alpha,\beta)$,  
and setting $\alpha=\beta=0$ in \eqref{eq:PMP} recovers the C-PMP condition for $\boldsymbol{\Sigma}(\boldsymbol{\theta})$.  
By Lemma~\ref{lemma:diff_of_xi}, for any $\boldsymbol{\theta}$ near $\bar{\boldsymbol{\theta}}$,  
$\boldsymbol{\xi}(\boldsymbol{\theta})=\tilde{\boldsymbol{\xi}}$ at $(\boldsymbol{\theta},\alpha,\beta)=(\boldsymbol{\theta},0,0)$ 
is a differentiable local isolated minimizing trajectory of $\boldsymbol{\Sigma}(\boldsymbol{\theta})$ with the associated unique \(C^1\) function.
Therefore, due to the uniqueness and once-continuous differentiability of $\tilde{\boldsymbol{\xi}}$ with respect to $(\boldsymbol{\theta},\alpha,\beta)$ near $(\bar{\boldsymbol{\theta}},0,0)$, one can obtain
$\boldsymbol{\xi}(\boldsymbol{\theta},\alpha,\beta)\rightarrow\boldsymbol{\xi}(\bar{\boldsymbol{\theta}},0,0)=\boldsymbol{\xi}(\boldsymbol{\theta}) \  \text{as} \  \alpha,\beta\rightarrow 0$
due to that $\boldsymbol{\xi}(\boldsymbol{\theta},\alpha,\beta)$ is unique and continuous at $(\boldsymbol{\theta},\alpha,\beta)=(\boldsymbol{\theta},0,0)$, and
$\frac{\partial \boldsymbol{\xi}(\boldsymbol{\theta},\alpha,\beta)}{\partial\boldsymbol{\theta}}\rightarrow\frac{\partial \boldsymbol{\xi}(\boldsymbol{\theta},0,0)}{\partial\boldsymbol{\theta}}=\frac{\partial\boldsymbol{\xi}(\boldsymbol{\theta})}{\partial\boldsymbol{\theta}} \  \text{as} \  \alpha,\beta\rightarrow 0$
due to that $\boldsymbol{\xi}(\boldsymbol{\theta},\alpha,\beta)$ is unique and once-continuously differentiable at $(\boldsymbol{\theta},\alpha,\beta)=(\boldsymbol{\theta},0,0)$.

\end{document}